\documentclass[conference]{IEEEtran}
\IEEEoverridecommandlockouts
\usepackage{cite}
\usepackage{amsmath,amssymb,amsfonts}
\usepackage{algorithmic}
\usepackage{graphicx}
\usepackage{textcomp}
\usepackage{xcolor}
\def\BibTeX{{\rm B\kern-.05em{\sc i\kern-.025em b}\kern-.08em
    T\kern-.1667em\lower.7ex\hbox{E}\kern-.125emX}}
\usepackage{adjustbox}
\usepackage{pifont}

\usepackage{multirow}
\usepackage[linesnumbered,ruled,vlined]{algorithm2e}
\usepackage{subcaption}
\usepackage{colortbl}
\usepackage{array}
\usepackage{url}
\newcolumntype{K}[1]{>{\centering\arraybackslash}p{#1}}

\begin{document}

\title{Energy-aware Distributed Microservice \\Request Placement at the Edge
\thanks{This work was supported by the Swedish national graduate
school in computer science (CUGS).}
}

\author{\IEEEauthorblockN{Klervie Toczé}
\IEEEauthorblockA{\textit{Dept. of Computer and Information Science} \\
\textit{Linköping University}\\
Linköping, Sweden \\
klervie.tocze@liu.se}
\and
\IEEEauthorblockN{Simin Nadjm-Tehrani}
\IEEEauthorblockA{\textit{Dept. of Computer and Information Science} \\
\textit{Linköping University}\\
Linköping, Sweden \\
simin.nadjm-tehrani@liu.se}
}

\maketitle

\begin{abstract}
  Microservice is a way of splitting the logic of an application into small blocks that can be run on different computing units and used by other applications. It has been successful for cloud applications and is now increasingly used for edge applications.  
  This new architecture brings many benefits 
  but it makes deciding where a given service request should be executed (i.e. its placement) more complex as every small block needed for the request has to be placed. 
  
  In this paper, we 
  investigate decentralized request placement (DRP) for services using the microservice architecture. We consider the DRP problem as an instance of a traveling purchaser problem and propose an integer linear programming formulation. This formulation aims at minimizing energy consumption while respecting latency requirements. 
  We consider two different energy consumption metrics, namely overall or marginal energy, to study how optimizing towards these impacts the request placement decision. 
  
Our simulations show that the request placement decision can indeed be influenced by the energy metric chosen, leading to different energy reduction strategies. 
\end{abstract}

\begin{IEEEkeywords}
Energy metrics, optimization, edge computing. 
\end{IEEEkeywords}

\section{Introduction}

The edge computing paradigm, consisting of moving computational and storage resources to the edge of the network, is envisioned to obtain  lower latencies, higher privacy and to alleviate the amount of data sent to a distant cloud. 
Thus, the next generation of wireless networks envisage deploying user connection and services in close vicinity of the end users. 
At the same time as edge computing infrastructures are being deployed, the microservice architecture is subject to intensive study, both in the cloud and at the edge. With this architecture, services are decomposed into a chain of different functions, allowing for higher flexibility and sharing of the function logic between different services. 

\begin{figure}
    \centering
    \includegraphics[width=0.8\columnwidth]{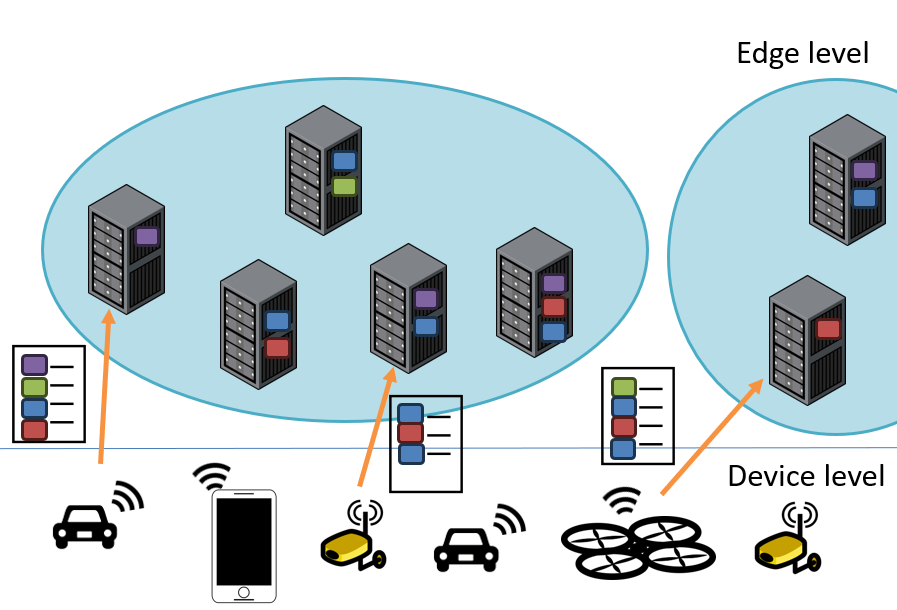}
    \caption{The considered edge system.}
    \label{fig:system}
\end{figure}

We consider a distributed edge infrastructure (as shown in Figure \ref{fig:system}), where each edge device receives service requests coming from end users through end devices. An end device can for example be an IoT device (a video surveillance camera, a sensor), a mobile phone, a connected vehicle, etc..
The edge computing infrastructure is composed of heterogeneous resource-limited devices. This means that functions cannot deploy function instances on all available edge devices as this will be too resource-hungry and also because some specific function instances can only run on specific edge hardware. For example, some machine learning algorithms require GPU resources to run. How to best deploy the function instances for improved performance and how many of them should be deployed are related placement problems but are tackled by other works~\cite{Tocze_Violinn, Rubak_PredictiveResourceScaling}. 

In addition to performance considerations, the energy use of edge computing systems has recently been attracting research interest. Whether edge computing can contribute to reducing the electricity consumption of ICT infrastructures and the associated carbon footprint has to be investigated~\cite{Varghese_RevisitingEdgeComputing}. Some efforts in this direction include works for estimating the energy consumption of edge infrastructure~\cite{Ahvar_EstimatingEnergy} and for using renewable energy sources~\cite{Perin_SustainableEdgeComputing}. 

In this work, we study request placement, i.e. for a given set of function instances already deployed in the edge infrastructure, which one(s) to select for executing a particular request, incoming at a given physical location and at a given point in time. We envisage that the decision about request placement is done locally, in a distributed way. Previous work~\cite{Schneider_DistributedOnline} has shown that distributed approaches with partial local knowledge can perform better than corresponding centralized ones.  Indeed, an edge environment is very dynamic, with user mobility causing arrival of service requests at varying rates 
and with different patterns according to events in the real world (e.g. gatherings, traffic jams, etc.). 
Hence, the edge load may vary drastically from one area to another over a short period of time, making it very complex to maintain global knowledge about the full system. With a distributed approach, such global knowledge is unnecessary, limiting the need for extensive message exchange that would occupy already scarce network resources. 

The request placement subproblem is critical to be addressed since it is where the demand side (the application requirements) and the supply side (the infrastructure provisioning) meet. 
The application provider is interested in the requests completing in time to guarantee a high quality of experience for its users. The edge infrastructure provider is interested because the request placement can be used to improve load balancing 
or  load consolidation. 
One especially critical moment where request placement is important is during transient system overloads (also known as flash crowds). 
These events are limited in time and spread in geographical location. Thus, it is not desirable 
to address them through over-provisioning, especially viewed from a resource minimization perspective.

In this paper, we formulate the decentralized request placement (DRP) problem as an instance of the Traveling Purchaser Problem. The corresponding optimization problem is solved using Integer Linear Programming (ILP). There are several optimization objectives that can be pursued in general and in edge computing in particular. A common one, as many edge applications are latency-critical, is to minimize the completion time of a service request. Instead of following this way, we argue that there is no benefit for minimizing the service completion time as long as it is below the deadline requirement, i.e. that its responsiveness is good enough. Therefore, we use the deadline as a constraint and choose to optimize with regards to energy consumption. 

We consider two different energy metrics having different underlying ideas: one considering the energy consumption as traditionally done and one considering the energy consumption increment created by the current request being placed. Our goal is to study how the choice of energy metric  can influence the optimal request placement decision and to which extent. 

The contributions of this paper are the following:
\begin{itemize}
    \item We express the DRP problem of a microservice request as a Traveling Purchaser Problem (TPP) and propose an ILP formulation for it.
    \item We formulate two energy metrics and use them in the ILP as two different optimization objectives. 
    \item We study the influence of varying load and change of energy metric 
    on the placement decision and evaluate the time overhead of the ILP approach. 
\end{itemize}

The rest of the paper is organized as follows: in Section \ref{sec:Notations} we introduce the system and the two considered energy metrics. Then, Section \ref{sec:Problem} presents the problem and its two formulations. 
The ILP formulation is implemented and evaluated as described in Section \ref{sec:Evaluation} with regards to energy consumption and timeliness. 
The outcomes are presented in Section \ref{sec:EvaluationOutcomes}, including a study of the influence of varying different system characteristics on the placement decision. We discuss related works in Section \ref{sec:RelatedWorks} and conclude in Section \ref{sec:Conclusion}. 

\section{System Model}
\label{sec:Notations}

This section presents the different concepts and models describing the considered edge system. It also introduces the two energy consumption metrics used in this work. 

\subsection{Edge infrastructure}

We consider that the whole edge infrastructure is divided into smaller areas, called orchestration areas, in order to enable higher locality-awareness in the placement decisions. One of these orchestration areas can for example represent a district/neighborhood of a city. In this paper, we limit our study to one orchestration area, served by several edge nodes.

The edge devices included in this orchestration area are represented as a directed graph\footnote{We choose not to use the classic E notation for edges to avoid confusion with the edge of edge computing. Instead we use an L for "Links". } $G=(V,L)$.
Each vertex of the graph corresponds to an edge device and each link of the graph corresponds to a communication link between two edge devices.

Within one orchestration area (i.e. one graph), we assume that every edge device has some information about the other edge devices and how they are connected.  
For example, regarding the edge devices, such information may include their current utilization level and their performance and energy profile. Information about the links between edge devices may include their propagation delay and transmission power. 
How this information is retrieved and how often this retrieval takes place is out of the scope of this paper. 

\subsection{Service}

The end users within one orchestration area want to use various \textit{services}. For example, an augmented reality application is used to visualize an architectural project, or the video footage of a crossing needs to be analyzed.

In this paper, the considered services are following the microservice architecture, i.e. "an approach for developing a single application as a suite of small services, each running in its own process and communicating with lightweight mechanisms"~\cite{Lewis_MicroserviceDef}. 
Therefore, a service $S$ can be defined as a directed acyclic graph $S=(F,D)$ where a vertex $f \in F$ is  one of the functions (micro services) the service is composed of and an edge $d \in D$ is a dataflow between two different functions~\cite{Li_ServerlessComputing_Survey}. 
A service $S$ is therefore a chain of functions connected with a sequence of edges in the same direction.
A function $f\in F$ is associated with a computing size $f^s$ and a dataflow $d \in D$ is associated with a data size $d^s$.  

In order to provide a service $S$, function instances for every function $f \in F$ are deployed on  different edge devices $v$ present in the infrastructure. At most one function instance for a given function $f$ is deployed on a given node $v$. We therefore introduce $\phi=(v,f)$ as a shorthand for describing the function instance for function $f$ deployed on node $v$. 
The function instances can then handle incoming requests. 

To execute a function which is a part of a service, an edge device needs to have the necessary hardware and software available. This is why all functions may not execute on all edge devices. The necessary software is for example provided using a dedicated container~\cite{Oleghe_ContainerPlacementAndMigration}. 
We denote $V_f \subseteq V$ the subset of edge devices able to execute a given function $f$, i.e. the subset of edge devices having a function instance for function $f$ deployed.  
We consider that each function is available on a subset of edge devices only ($\forall f, V_f \subset V$) and that every function is available on at least one edge device ($\forall f, V_f\neq \emptyset$).

\subsection{Service request}

The load coming to the edge infrastructure consists of service requests. A service request $r=(S,v_b,v_e,\theta, \delta) \in R$ is defined by:
\begin{itemize}
    \item the requested service $S$
    \item the edge device receiving the request (beginning device) $v_b \in V$
    \item the edge device receiving the request answer (end device) $v_e \in V$ 
       \item the request arrival time $\theta$ 
    \item the request deadline $\delta$
\end{itemize}

The deadline $\delta$ is relative to the service request arrival time $\theta$ and corresponds to the maximum time allowed for the request to go through all the functions comprising the service and reach the destination device.  
In this work, we consider services for which it is necessary to complete requests before the deadline, otherwise their quality of service (QoS) is severely degraded.

\subsection{Edge orchestration}
\label{sec:EdgeOrchestration}

The placement problem within an edge orchestration area can be decomposed into three different subproblems. 

The \textbf{request placement subproblem}: for each request $r$ with service $S$ incoming to the edge system, the edge system has to decide on  a placement $p$ so that the request $r$ will be completed before its deadline $\delta$. 

The \textbf{service placement subproblem}: only a subset of the edge devices contain the required environment to run a service at a given point in time (i.e. the service instances). This avoids reserving resources for a service which is not used by users in this area at the moment. The subproblem consists in determining this subset after every major load change such that the incoming load is served with as high QoS as possible. 

The \textbf{edge device provisioning subproblem}: sometimes, the incoming load to the system can only be handled by increasing the amount of resources (namely the edge computing devices) available at a certain location. Therefore, at every load change requiring it, this subproblem consists of enabling minimum extra resources at specific locations such that the incoming load is served with as high QoS as possible. 

In this paper, we focus on the request placement subproblem. We therefore assume that 1) the edge devices have been provisioned and that no extra edge resource is available, and 2) a set of function instances has been placed on various edge devices in order to be able to provide for the services asked by the end users. Both the edge device provisioning and the service 
 are fixed in the context of this work. 
The reader is referred to other works where a combination of these problems are addressed using heuristics~\cite{Tocze_Violinn,Tocze_ORCH}. 

\subsection{Request placement}

Deciding on which function instance (and by extension on which edge device) each of the functions of a service $S$ will be executed when handling a given service request is called placing the request. Such a service request placement is denoted $p=(\phi_1,...,\phi_{|F|})$ where $\phi_i$ corresponds to the function instance chosen for executing the $i^{th}$ function of service $S$. 

From a given placement $p$,  the set of edge devices included in the placement $V_p\subseteq V$ (i.e. the ones where the selected function instances are deployed) is therefore $V_p = \{v | \phi=(v,f) \in p\}$. $L_p\subseteq L$ is the set of links connecting theses edge devices.

In this work, we focus on 
request placement taking place at a given instance with a specific current resource utilization.  
 The aim is to be able to study how different optimization objectives (in particular regarding energy consumption) will influence the placement decision.

\subsection{Request completion time}
\label{sec:requestCompletionTime}
Once the placement $p$ of a service request $r$ is known, it is possible to calculate the completion time of the request in order to see whether it is below the deadline $\delta$. 

The completion time of a request is composed of two parts:
\begin{itemize}
    \item The transmission time
    \item The execution time.
\end{itemize}

The transmission time depends on the link propagation delay (in ms), the available link bandwidth (in byte/ms), and the size of the data (in byte) that needs to be transferred on the link. 
We assume that the request utilizes all the available link bandwidth to transmit. Therefore, the transmission time of dataflow $d$ with data size $d^s$ on a given link $l$ is:
\begin{equation}
\label{eq:linkCompletionTime}
    \lambda_{ld}=l^l+\frac{d^s}{l^c}
\end{equation}
where $l_l$ is the propagation delay for link $l$, and $l^c$ is the currently available bandwidth for the link $l$. 
We assume that the link(s) chosen correspond to the shortest path between the two edge devices.

The execution time depends on the available computing capacity of the edge device (in million of instruction (MI) per millisecond) and the size of the function needing to be computed (in MI). 
Each function instance running on the edge device gets a share of the full capacity. This share (i.e. the available computing capacity for the function instance) can account e.g. for the need to always have some free capacity and thus may vary over time.
We assume that the request utilizes all the available computing capacity in the corresponding function instance for the function execution. 
The execution time of the function instance $\phi=(v,f)$ corresponding to function $f$ with computing size $f^s$ deployed on edge device $v$ is therefore calculated as follows:
\begin{equation}
\label{eq:sys_executionTime}
     \lambda_{\phi}=\frac{f^s}{\phi^c}
\end{equation}
 where $\phi^c$ is the available computing capacity allocated to the function instance $\phi$. 

The total completion time of a service request $r$ to service $S=(F,D)$ using a given placement $p$ is the sum  of the transmission times and execution times for all link transfers and function executions required to complete the service request.
This can be expressed as:
\begin{equation}
\begin{split}
    \label{eq:latencyTotal}
    \Lambda_{rp}=\sum_{l \in L_p}\sum_{d \in D}\mathcal{I}^{p}_{ld}\lambda_{ld}+\sum_{v \in V_p}\sum_{f \in F}\mathcal{I}^{p}_{vf}\lambda_{\phi} \\
    \text{where} \qquad \mathcal{I}^{p}_{ld}\begin{cases}
         1, & \text{if $d$ is sent over $l$ according to $p$}\\
            0, & \text{otherwise.}
            \end{cases}
    \\
    \text{and} \qquad \mathcal{I}^{p}_{vf}\begin{cases}
         1, & \text{if $f$ executes on $v$ according to $p$}\\
           0, & \text{otherwise.}
            \end{cases}
\end{split}
\end{equation}

\subsection{Request energy consumption}
\label{sec:EnergyConsumption}

A major concern in placement decisions is the energy consumption associated to the service placement. In this paper, the aim is to strive for low resource use for every request placement. 
Since using computing devices and communication links is associated with some energy consumption, we use the utilization of links and computing devices as a proxy for measure of energy use. 

The request energy consumption, in a similar way to the request completion time, has two parts: 
\begin{itemize}
    \item The transmission energy consumption
    \item The execution energy consumption.
\end{itemize}

Following Baccarelli et al.~\cite{Baccarelli_EcoMobiFog} and Ahvar et al.~\cite{Ahvar_EstimatingEnergy}, we model the energy consumed by an edge device or an edge link as having both a static part (corresponding to the device/link being in the idle state) and a dynamic part (corresponding to the energy needed for processing/transmitting).

The energy used by an edge link $l$  to service a request through the transmission of the dataflow $d$ is calculated as follows:
\begin{equation}
\label{eq:energyLink}
    E_{ld}=\underbrace{\mathcal{P}^{IDLE}_l*\lambda_{ld}}_{\text{static part}}+\underbrace{\mathcal{P}^{DYN}_l*\lambda_{ld}}_{\text{dynamic part}}
\end{equation}
where $\mathcal{P}^{IDLE}_l$  is the idle power needed for maintaining the link $l$ (e.g. the power consumed by the NIC cards at both ends when in the idle state) and $\mathcal{P}^{DYN}_l$ is the sum of the power needed by the link $l$ for transmitting from the transmitting node and receiving at the receiving node. These are related to  a wide range of characteristics of the communication link (e.g. number of antennas) and the current link throughput~\cite{Baccarelli_EcoMobiFog}. $\lambda_{ld}$ is the duration of using $l$ for transmitting $d$, according to Section \ref{sec:requestCompletionTime}. 

For the function instances running on the edge devices, the power for the dynamic part is modeled as a linear relationship between power and utilization level of the function instance, assuming that one core is allocated to every function instance~\cite{Ahvar_EstimatingEnergy}. 
The adaptation of Equation \ref{eq:energyLink} to different functions in terms of relation of energy to dynamic power and computation time is possible, but left out in this paper. 

We introduce two different energy metrics for using a function instance $\phi$:
\begin{enumerate}
    \item The overall energy consumption, i.e. how much the device will consume after placing the execution of $f$ on top of what it is already executing.
    \item The marginal energy consumption, i.e. how much additional energy does the execution of function $f$ consume on that device.
\end{enumerate}

The overall energy consumption for an edge device $v$ used to service a request  through the execution of the function $f$ is therefore:
\begin{equation}
    E^O_{\phi}=\underbrace{\mathcal{P}^{IDLE}_v*\lambda_{\phi}}_{\text{static part}}+\underbrace{\mathcal{P}^{DYN}_v(u_{\phi})*\lambda_{\phi}}_{\text{dynamic part}}
\end{equation}
where  $\mathcal{P}^{IDLE}_v$ is the power needed for device $v$ to be on and $\mathcal{P}^{DYN}_v(u_{\phi})=(\mathcal{P}^{100\%}_v-\mathcal{P}^{IDLE}_v)*u_{\phi}$ is the extra power needed for device $v$ for executing at a utilization level of $u_{\phi}$ (that includes the execution of the function instance $\phi$). $\lambda_{\phi}$ is the duration of executing $f$  on $v$, according to Section \ref{sec:requestCompletionTime}.
For the overall energy metric, we consider $\mathcal{P}^{IDLE}_v$, i.e. the idle power needed for the full edge device to be on. This corresponds to the worst case where only a function instance for function $f$ is deployed on edge device $v$. If several function instances are deployed, one can define variants of this metric where only a part of the edge device idle power is taken into account for the energy consumption of a given instance. How to determine this part can be done using energy apportionment~\cite{Vergara_Apportionment}. Such variants are out of the scope of this work.   

The marginal energy consumption for an edge device $v$ used to service a request  through the execution of the function $f$ depends on what was the utilization of device $v$ before the execution of $f$ ($u_v$). It is calculated as:
\begin{equation}
\label{eq:energyDevice}
    E^M_{\phi}=\begin{cases}
         E^O_{\phi}, & \text{if $u_v =0$ }\\
            \mathcal{P}^{DYN}_v(u_{\phi}-u_{v})*\lambda_{\phi}, & \text{otherwise.}
    \end{cases}
\end{equation}

The difference between the overall and marginal energy consumption for a given execution of a function instance represents whether we need to use a previously unused device for the execution (first line in Equation \ref{eq:energyDevice}) or whether the function is added to an already used device. This is useful to know because if the static energy consumption represents an important part of the energy consumption, then a placement should favor the devices already in use instead of needing to turn on new ones. 

Finally, the total overall energy consumption of a service request $r$ to service $S=(F,D)$ using a given placement $p$ is obtained by summing the corresponding energy consumption of all dataflow transmissions and the corresponding energy consumption for computing each function on the edge devices included in the placement solution. 
This is expressed for the overall energy consumption as:
\begin{equation}
\begin{split}
    \label{eq:energyTotal}
    E^O_{rp}=\sum_{l \in L_p}\sum_{d \in D}\mathcal{I}^{p}_{ld}E_{ld}+\sum_{v \in V_p}\sum_{f \in F}\mathcal{I}^{p}_{vf}E^O_{\phi}
    \\
    \text{where} \qquad \mathcal{I}^{p}_{ld}\begin{cases}
         1, & \text{if $d$ is sent over $l$ according to $p$}\\
            0, & \text{otherwise.}
            \end{cases}
    \\
    \text{and} \qquad \mathcal{I}^{p}_{vf}\begin{cases}
         1, & \text{if $f$ executes on $v$ according to $p$}\\
           0, & \text{otherwise.}
            \end{cases}
    \end{split}
\end{equation}
A similar formula is obtained for the marginal energy consumption, including the terms $E^M_{rp}$ and $E^M_{\phi}$ instead. 

\section{DRP Formulations} 
\label{sec:Problem}

In this work, the request placement subproblem from Section \ref{sec:EdgeOrchestration} is solved in a decentralized manner. It means that the requests arrive to the closest edge device in the orchestration area and it is this receiving edge device that is responsible for deciding on the placement, i.e. deciding which function instances to select on which other devices to complete a request. 

\subsection{TPP formulation}
The service request placement subproblem (i.e. the DRP problem) can be expressed as an instance of the Traveling Purchaser Problem (TPP)~\cite{Manerba_TPPandVariants}. 

The TPP is a generalization of the well-known Traveling Salesman Problem (TSP), where there are different marketplaces that sell a given set of items at a given price. The problem is defined for a purchaser that has a given list of items to buy, to find the route between the marketplaces that minimizes both the cost of travel and purchase. 

In our case, the purchaser corresponds to the service request that has to travel to different edge devices (marketplaces) offering the functions (items) composing the service requested (the list of items). 
The approach is energy-aware and considers the costs of travel and purchase to be the energy consumption of using a link or an edge device, according to their descriptions in Section \ref{sec:EnergyConsumption}. 

In order to adopt this solution, we need to add two constraints to the generic formulation of the TPP. The first one is that there is a given order in which the functions should be executed, so that the service is executed properly. Secondly, it is not enough to minimize the cost of the request placement, it should also meet its deadline to be an acceptable solution.

\subsection{ILP formulation}
An optimal solution to the service request placement problem can be found using integer linear programming (ILP). 
Table \ref{tab:notationILP} summarizes the notations used for the formulation presented below. 

\begin{table}[]
    \centering
    \begin{tabular}{|K{2cm}|p{0.7\columnwidth}|}
    \hline
        \textbf{Symbol} & \textbf{Meaning} \\
        \hline
       
        $x_{\phi\psi}$& Decision variable indicating whether the link between function instances $\phi$ and $\psi$ is included in the solution\\\hline
        $y_{\phi}$& Decision variable indicating whether  the function instance $\phi$ is included in the solution \\\hline
        $o_{\phi\psi}$& Decision variable indicating the order in which the link between function instances $\phi$ and $\psi$ is visited in the solution \\\hline
        $\Phi 
        $& Set of function instances\\\hline
                  $F= (f_1,f_2,....,f_{|F|})$& Ordered list of service functions to be placed\\\hline 
         $\phi_b$& Beginning function instance located on device $v_b$\\\hline
         $\phi_e$& End function instance located on device $v_e$\\\hline
         $E_{(\phi,\psi)}$& Energy consumption of the link between any two function instances $\phi$ and $\psi$ (including zero for modelling both being on the same device)\\\hline
         $E_{\phi}$& Energy consumption of the function instance $\phi$ \\\hline
         $a^{\phi}_f$& Boolean indicating whether function instance $\phi$ is an instance of function $f$ (=1) or not (=0). \\\hline
         $O_{f}=i$& Positive integer representing the position of the function $f$ in the list $F$ \\\hline
         $\lambda_{\phi\psi}$& Transmission time for the link between the function instances $\phi$ and $\psi$\\\hline
         $\lambda_{\phi}$& Execution time for the function instance $\phi$\\\hline
         $\beta$& Integer upper bound used when ordering the visited links. It should be greater that the number of links contained in the service chain. 
         \\\hline
    \end{tabular}
    \caption{ILP notations.}
    \label{tab:notationILP}
\end{table}

\subsubsection{The infrastructure graph}

In order to enforce the ordering constraint between the different functions composing the service, we use a variable assigning an integer to the links that are part of the placement (in a similar way to the work by Shameli-Sendi et al.~\cite{Shameli-Sendi_EfficientProvisioning}). In order to both enable a placement to have several service instances on the same physical node (good for load consolidation) and to allow the same edge device to be selected for different (non-consecutive) functions, 
we consider a device in the ILP infrastructure graph to be  a function instance and not a hardware device. Consequently, a link is a virtual link between function instances, that can be mapped to a physical one.
We also represent the beginning and end devices ($v_b$ and $v_e$) as two specific "virtual" function instances denoted $\phi_b$ and $\phi_e$ respectively.  

Note that since the nodes in the ILP graph are function instances, it is not necessary to specify both the edge device and the function (as a function instance is the combination of both) where referring to a node. This means that for the completion time and energy consumption equations (Equations \ref{eq:linkCompletionTime} to \ref{eq:energyTotal}), the subscript $ld$ can be replaced by the corresponding start and end function instances. 

\subsubsection{Decision variables}

This formulation uses three different decision variables. 
$x_{\phi\psi}$ indicates whether the link between function instances $\phi$ and $\psi$ is selected to be part of the solution.
$y_{\phi}$ indicates whether the function instance $\phi$ is included in the placement. 
$o_{\phi\psi}$ captures the order of visiting the links in the solution. If the link $(\phi,\psi)$ is not visited, then $o_{\phi\psi}=0$. If the link $(\phi,\psi)$ is visited before the link $(\chi,\zeta)$ then we have  $o_{\phi\psi} > o_{\chi\zeta}$.
In case the link $(\phi,\psi)$ is visited just before the link $(\chi,\zeta)$ then we have  $o_{\phi\psi} = o_{\chi\zeta}+1$.

\subsubsection{Objective function}
\label{sec:objectiveFunction}

The objective function in this work is focusing on energy consumption as a proxy for resource consumption in the broader sense. 
The energy consumption metric considered here is either the overall or the marginal one described in Section \ref{sec:EnergyConsumption}. The idea is that optimizing against the marginal energy consumption metric favours a placement using load consolidation while the overall energy consumption metric, which is the one traditionally used, may not.  
This work is focusing on minimizing resource use, therefore the objective function is: 
$$\text{Minimize } \sum_{\phi \in \Phi}\sum_{\psi \in \Phi} E_{\phi\psi}* x_{\phi\psi} + \sum_{\phi \in \Phi\setminus\{\phi_b,\phi_e\}} E_{\phi}*y_{\phi}$$

The energy consumption metric $E$ in the equation above is replaced by the overall ($E^O$) or the marginal ($E^M$) energy consumption metric depending on the one chosen for optimizing. 

\subsubsection{Constraints}
The ILP formulation contains the following constraints:
\begin{equation}
    \label{eq:S_edge}
    \sum_{\psi \in \Phi} x_{\psi\phi} = \sum_{\zeta \in \Phi} x_{\phi\zeta} \qquad\forall \phi \in \Phi\setminus\{\phi_b,\phi_e\}
\end{equation}
\begin{equation}
    \label{eq:S_edge_start}
    \sum_{\psi \in \Phi} x_{\psi\phi} + 1 = \sum_{\zeta \in \Phi} x_{\phi\zeta} \qquad  \phi = \phi_b
\end{equation}
\begin{equation}
    \label{eq:S_edge_final}
    \sum_{\psi \in \Phi} x_{\psi\phi}  = \sum_{\zeta \in \Phi} x_{\phi\zeta} + 1 \qquad \phi = \phi_e
\end{equation}

\begin{equation}
    \label{eq:S_quantity}
     \sum_{\phi \in \Phi\setminus\{\phi_b,\phi_e\}} (y_{\phi}*a_f^{\phi}) = 1 \qquad\forall f \in F
\end{equation}
\begin{equation}
 \label{eq:S_include}
     \sum_{\psi \in \Phi} x_{\phi\psi} -y_{\phi} = 0  \qquad\forall \phi \in \Phi\setminus\{\phi_b,\phi_e\} 
\end{equation}

\begin{equation}
 \label{eq:S_avoid1}
    \sum_{\phi \in \Phi} x_{\phi\psi} \leq 1  \qquad \forall \psi \in \Phi
\end{equation}
\begin{equation}
 \label{eq:S_avoid2}
    \sum_{\psi \in \Phi} x_{\phi\psi} \leq 1  \qquad \forall \phi \in \Phi
\end{equation}

\begin{equation}
 \label{eq:S_visit}
    o_{\phi\psi} \leq \beta * x_{\phi\psi}  \qquad \forall \phi \in \Phi,\forall \psi \in \Phi
\end{equation}
\begin{equation}
 \label{eq:S_calcul}
\sum_{\psi \in \Phi}o_{\psi\phi} = \sum_{\psi \in \Phi}(o_{\phi\psi} +x_{\phi\psi}) \qquad \forall \phi \in \Phi \setminus \{\phi_b,\phi_e\}
\end{equation}
\begin{equation}
 \label{eq:S_calcul_source}
\sum_{\psi \in \Phi}o_{\psi\phi} + \beta = \sum_{\psi \in \Phi}(o_{\phi\psi} +x_{\phi\psi}) \qquad \phi =\phi_b 
\end{equation}

\begin{equation}
 \label{eq:S_order}
 \begin{aligned}
     (\sum_{\alpha\in \Phi\setminus \{\phi_e\}} o_{\alpha\phi}) -x_{\phi\psi} - \beta*y_{\phi} +\beta \geq \\ (\sum_{\alpha\in \Phi\setminus \{\phi_e\}} o_{\alpha\psi}) - \beta * y_{\psi} +\beta * \sum_{\alpha\in \Phi\setminus \{\phi_e\}} x_{\alpha\psi}  \\ 
     \forall \phi,\psi \in \Phi \setminus \{\phi_b,\phi_e\}, \phi \ne \psi, \\ \forall f,k \in F, f \ne k, O_{f} = O_{k} -1, a^{\phi}_f=a^{\psi}_k=1
 \end{aligned}
\end{equation}

\begin{equation}
 \label{eq:S_deadline}
    \sum_{\phi \in \Phi}\sum_{\psi \in \Phi}\lambda_{\phi\psi}*x_{\phi\psi}+\sum_{\phi \in \Phi \setminus \{\phi_b,\phi_e\}}\lambda_{\phi}*y_{\phi} \leq \delta \qquad  
\end{equation}

\begin{equation}
 \label{eq:S_noSelfLoop}
    \sum_{\phi \in \Phi}x_{\phi\phi}=0  
\end{equation}

\begin{equation}
 \label{eq:S_binX}
     x_{\phi\psi} \in \{0,1\}  \qquad\forall \phi,\psi \in \Phi
\end{equation}
\begin{equation}
 \label{eq:S_binY}
     y_{\phi} \in \{0,1\}  \qquad\forall \phi \in \Phi \setminus \{\phi_b,\phi_e\}
\end{equation}
\begin{equation}
 \label{eq:S_binV}
    o_{\phi\psi}\in \mathbb{N}, \qquad 0 \leq o_{\phi\psi} \leq \beta,    \qquad\forall \phi,\psi \in \Phi 
\end{equation}

Constraint \ref{eq:S_edge} ensures that for any function instance $\phi$, there is one edge going in and out the function instance. For the special cases of the source (Constraint \ref{eq:S_edge_start}) and destination (Constraint \ref{eq:S_edge_final}), a virtual incoming (respectively outgoing) edge is added to them.
Constraint \ref{eq:S_quantity} ensures that only one  function instance is selected per function to be placed and that the function instance selected can actually run this function. 
Constraint  \ref{eq:S_include} ensures that if a function instance is chosen to be included in the solution, then it has to be on the solution path. 
Constraints  \ref{eq:S_avoid1} and \ref{eq:S_avoid2} ensure the solution does not include any cycle. This is needed since the ordering constraint can only assign one order number to each link, hence cycles are impossible. 
Constraint  \ref{eq:S_visit} ensures that if a link is not selected, the value of the corresponding $o_{\phi\psi}$ is set to 0. If a link is selected, the corresponding ordering variable $o_{\phi\psi}$ is less than $\beta$. 
Constraint  \ref{eq:S_calcul} is used to calculate the values of $o_{\phi\psi}$ for each selected link.
The special case of the start node is covered in Constraint \ref{eq:S_calcul_source}.
Constraint  \ref{eq:S_order} ensures that the functions are placed in the indicated order. The order is specified by $O$, where if $O_l = O_{k}-1$, it means that the function $l$ is placed before function $k$ and that the function instance $\phi$ where $y_{\phi}*a^{\phi}_l=1$ is visited before the function instance $\psi$ where $y_{\psi}*a^{\psi}_k=1$. 
Constraint \ref{eq:S_deadline} ensures that the latency corresponding to the selected path (i.e. a set of function instances connected through ordered links) is below the latency requirement, i.e. the deadline of the service request.
Constraints \ref{eq:S_binX} and \ref{eq:S_binY} indicate that the decision variables are binary. Constraint \ref{eq:S_binV} indicates that the decision variable is a positive integer. 

\section{Evaluation setup}
\label{sec:Evaluation}

The proposed ILP formulation is implemented and this section describes how the solutions found are evaluated. 
Section \ref{sec:eval_edgeSystem} presents the characteristics of the system in which the placement of service requests are optimized. 
Next, the service that is evaluated and the load scenarios are described. Finally, the performance metrics that are gathered and the evaluation approach are presented.  

\subsection{Edge system}
\label{sec:eval_edgeSystem}
The edge system considered in this evaluation has the same topology as the Abilene network~\cite{Abilene}. The distances between the edge devices are scaled to correspond to a scenario where all the edge devices are spread within a neighborhood of a city. The propagation delay between two edge devices is taken as proportional to the distance between these two devices. This model can easily be adapted to consider different delay characteristics (e.g. link propagation depending on connection technology). 

A service request arrives at the closest edge device and is placed by this receiving edge device. Each edge device has some knowledge about the other edge devices located in its vicinity, which are the candidates for service request placement. Example of information is which function instances are available and current utilization levels. How this information is obtained and updated (and how accurate it is) is out of the scope of this paper. 

All the edge devices and links in the system have the same characteristics with regards to processing/transmitting capacity, as well as the same energy profile. A summary of the system characteristics is shown in Table \ref{tab:system}.  The energy values are taken from the measurements performed by Ahvar et al.~\cite{Ahvar_EstimatingEnergy} 

\begin{table}[]
    \centering
    \begin{tabular}{|c|c|}
    \hline
        \textbf{Characteristic} & \textbf{Value}  \\
        \hline
         Edge device computational capacity& 500 MI/ms \\
        \hline
        Edge device idle power& 98 W \\ 
        \hline
        Edge device dynamic power& 50 W \\ 
        \hline
        Link bandwidth & 500 MB/ms \\
        \hline
        Link idle power &  1 W\\ 
        \hline
        Link dynamic power &  9 W\\ 
        \hline
               
    \end{tabular}
    \caption{System characteristics}
    \label{tab:system}
\end{table}

\subsection{Service} 
\begin{figure*}
    \centering
    \includegraphics[width=\textwidth]{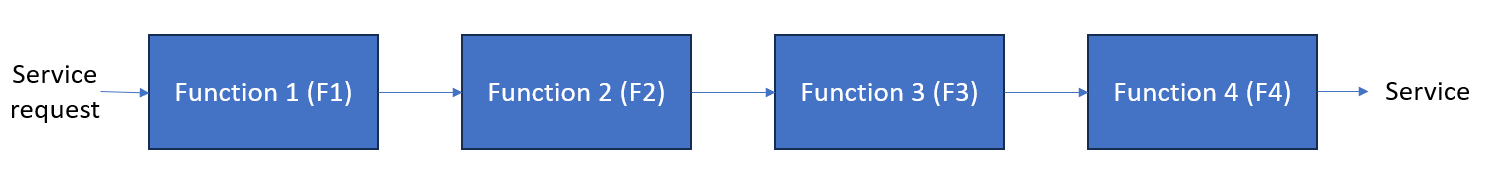}
    \caption{The edge service considered for evaluation}
    \label{fig:application}
\end{figure*}

The edge service considered in this evaluation is composed of four functions that are executed sequentially. It is illustrated in Figure \ref{fig:application}. Table \ref{tab:application} summarizes its different characteristics.

Moreover, we assume that the end device issuing the request waits for the result of the service request, meaning that the request is received at an edge device and the execution result have to be transmitted to this same edge device after the service request execution is completed. 
An example of such a service is a mixed reality application where service requests contain images that have to be decoded (F1), analyzed to create a virtual representation of the scene (F2), modified with the addition of virtual elements (F3) and encoded (F4), before being sent back to the issuing end device for rendering to the end user~\cite{Tocze_JoCC}.

\begin{table}[]
    \centering
    \begin{tabular}{|c|c|c|}
    \hline
        \textbf{Characteristic} & \multicolumn{2}{c|}{\textbf{Value}}  \\
        \hline
         Request deadline& \multicolumn{2}{c|}{100 ms} \\
        \hline
        \multirow{4}{*}{Computation} & Function 1 & 20 MI \\
        \cline{2-3}
        & Function 2 & 200 MI \\
        \cline{2-3}
        & Function 3 & 200 MI \\
        \cline{2-3}
        & Function 4 & 20 MI \\
        \hline
        \multirow{5}{*}{Communication} & Dataflow 1 & 250 MB \\
        \cline{2-3}
        & Dataflow 2 &  500 MB\\
        \cline{2-3}
        & Dataflow 3 & 750 MB \\
        \cline{2-3}
        & Dataflow 4 & 500 MB \\
         \cline{2-3}
        & Dataflow 5 & 250 MB\\
        \hline
        
    \end{tabular}
    \caption{Service characteristics}
    \label{tab:application}
\end{table}

Moreover, the number of function instances for each function and their localization is predefined. We recall that where to place them and how many of them should be placed are the focus of the other two subproblems of edge orchestration (see Section \ref{sec:EdgeOrchestration}) and thus out of the scope of this paper. In this evaluation, two function instances are deployed for each of the functions on two different edge devices and the edge devices chosen are different for the different functions.  

\subsection{Load scenario}
\label{sec:loadScenario}

The load in the evaluation consists of different utilization levels for the edge devices and communication links. These utilization levels indicate which share of the total computation/communication capacity is available for the request being placed. 
For this evaluation, the utilization levels (in percentage) are decided randomly using a Normal distribution. As a baseline we take a distribution with a mean of 50 and a standard deviation of 10. This represents a system that is neither highly overloaded nor severely underloaded.

\subsection{Performance metrics}
\label{sec:Evaluation_PerformanceMetrics}

The request placement solution is evaluated with regards to three different metrics.
The first metric is the request completion time. Even though a request is always required to  complete before its deadline, it can complete more or less fast depending on the optimization objective. 
The other two metrics measure energy consumption connected to the request. We first measure the overall energy consumption and then the marginal energy consumption, as defined in Section \ref{sec:EnergyConsumption}.

This evaluation uses different statistical measures such as mean, median, standard deviation, 10th and 90th percentile to present the different values of the above metrics in the different random load scenarios.  

\subsection{Evaluation approach}
\label{sec:Simulator}

The evaluation consists of groups of 25 runs each. Each run corresponds to a different random load scenario (as described in Section \ref{sec:loadScenario}). 
A group corresponds to a given load distribution and a given number of instances available for each function. A summary of the six different group types considered is presented in Table \ref{tab:evaluation_groups}. 

The columns present the characteristics of each group type, starting with an identifier (column \#) and a short name (column Name). The next two columns describe how the load for the edge devices, respectively the communication links is generated (columns load distribution edge devices/communication links). Then the number of function instances available for each function in the evaluation is shown (column \# instances per function). The last column indicates where the results from the given group types are presented in this paper. 

The rows present the different group types. Row 1 (Baseline) provides the scenario for an initial evaluation of the obtained placement solutions. Rows 2 to 4 are studying different load distributions for the edge devices. In rows 2 and 4, the load level on each edge devices is obtained using a normal distribution, with a standard deviation of 10 in row 2 and 30 in row 4. In row 3, the load level is the same for all edge devices and is fixed to a value between 0 and 100, increasing by steps of 10. Rows 5 and 6 are studying the impact of having more function instances available (4 for row 5 and 6 for row 6). The other parameters are the same as row 2. 

When the load distribution edge devices column contains a cell with the mention "varying-load-0-100", it means that the evaluation is repeated with the load level varying between 0 and 100 in steps of 10, the other parameters being the same. Hence, the corresponding row of the table contains 11 groups of runs, one for each load level value between 0 and 100.

\begin{table*}[]
    \centering
    \begin{tabular}{|c|c|K{5cm}|K{3.25cm}|K{2.25cm}|c|}
    \hline
         \textbf{\#}&\textbf{Name}&\textbf{Load distribution edge devices}&\textbf{Load distribution communication links} &\textbf{\# instances per function}& \textbf{Presented in} \\ \hline
        1&Baseline& $\mathcal{N}$(50, 10) &$\mathcal{N}$(50, 10) &2& Section \ref{sec:results} \\ \hline
        2&Normally distributed load& $\mathcal{N}$(varying-load-0-100, 10) &No load &2& Section \ref{sec:influenceLoadLevel}\\\hline
        3&Fixed load& Fixed load level at a value between 0-100 &No load&2& Section \ref{sec:influenceLoadLevel} \\\hline
        4&Larger standard deviation& $\mathcal{N}$(varying-load-0-100, 30)&No load&2& Section \ref{sec:influenceLoadLevel} \\\hline
        5&4 function instances&$\mathcal{N}$(varying-load-0-100, 10) &No load&4& Section \ref{sec:InfluenceAvailableInstances}\\\hline
        6&6 function instances&$\mathcal{N}$(varying-load-0-100, 10) &No load&6&Section \ref{sec:InfluenceAvailableInstances} \\\hline
    \end{tabular}
    \caption{Summary of the performed run group types.}
    \label{tab:evaluation_groups}
\end{table*}

In all runs of all groups, we place one request using our ILP formulation, both when minimizing overall energy consumption and also marginal energy consumption. These are named Overall and Marginal for short. In total there are 2800 runs executed. 

The evaluation is performed on a Dell Precision 5520 equipped with an Intel Xeon E3-1505M CPU (2.8 GHz) and 16GB of RAM. Version 9.0.2 of the Gurobi Optimizer 
is used as the optimization solver. 

The code used for running the evaluation as well as the result files are provided open-source in Gitlab\footnote{\url{https://gitlab.liu.se/ida-rtslab/public-code/2024_energy-aware_placement}}. 

\section{Evaluation outcomes}
\label{sec:EvaluationOutcomes}
This section presents and discusses the outcomes of the evaluation presented in Section \ref{sec:Evaluation}. First, the impact of each energy metric in the baseline runs is presented in Section \ref{sec:results}. Then, the impact of load level, load level heterogeneity and availability of function instances are further studied in Sections \ref{sec:influenceLoadLevel} and \ref{sec:InfluenceAvailableInstances}. Next, the impact on request completion time is discussed in Section \ref{sec:influenceCharacteristics}. Finally, a study of the time taken for solving the DRP problem is presented in Section \ref{sec:eval_timeToSolve}.

\subsection{Impact of each energy metric}
\label{sec:results}

In this section the baseline group is considered. It means that both edge devices and communication links utilization levels are randomly chosen with a Normal distribution with mean 50 and standard deviation 10. As all groups, it contains 25 runs. 
We will first show a detailed analysis of the outcomes obtained from one of the 25 runs for a given request. We will then 
describe the outcomes of the entire set of 25 runs.

For choosing the specific run, we select the first of the 25 runs where the placement solution differs with respect to the two optimization objectives, which was the second run of the group. Table \ref{tab:results} summarizes the results obtained for this specific run. The aim is to identify potential factors influencing the two 
objectives leading to different placement solutions. 

\begin{table}[]
    \centering
    \begin{tabular}{|l|c|c|}
    \hline
        Optimization objective & Overall energy & Marginal energy \\
        \hline
        Request completion time  & 85 ms      &  78 ms \\
        \hline
        Overall energy consumed  & 4595 J      & 4672 J  \\
        \hline
        Marginal energy consumed  & 1695 J      & 1623  J \\
        \hline
        Solver execution time & 64 ms   & 61 ms \\
        \hline
    \end{tabular}
    \caption{Evaluation results for one request placement.}
    \label{tab:results}
\end{table}

First, the performance of the different optimization objectives considered is analyzed with regards to the achieved request completion time. 
In both cases, it is possible to place the request so that it completes before its deadline. The Marginal objective results in a completion time faster by 8\% (corresponding to 7 ms). In both cases, there is at least 15 ms left before the deadline. 

Considering row 3 in Table \ref{tab:results}, we see that the strategy of optimising with regards to overall energy (column 2) leads to an actual overall energy consumed at a higher rate than the one achieved with the Marginal energy objective (column 3), leading to a 77J higher energy in row 3. 
Logically, the marginal energy (row 4) is lower for the Marginal energy objective (column 3) than for the Overall one (column 2). Moreover, the marginal energy amounts to 35 \% of the overall energy for Marginal and 37\% for Overall (comparing figures in rows 3 and 4).
Regarding the time needed for the solver to execute (row 5), it took a bit more than 60 ms for both 
objectives. 

As seen in Table \ref{tab:results}, the different optimization objectives lead to different overall and marginal energy values. To understand the actual decision better, the placement solution obtained is analyzed from the logs to understand how this difference appears. Depending on the optimization objective, the request is placed on different edge devices. Hence the two different energy consumption metrics (and of minimizing according to it) can actually result in different placements for the same request. The placement differs in the following way: functions 1, 2, and 4 are placed in the same function instances for both objectives, only the placement of function 3 is different. The difference in energy consumption shown in Table \ref{tab:results}, which is non-negligible, therefore it appears that even if only one of the four functions was placed in a different way a significant enough gain can be made.

In the logs, we look deeper at the placement, and especially the placement of F3. It is placed on device 6 in the Overall case, and on device 8 in the Marginal case. 
For the scenario considered, the devices 6 and 8 differ in the following ways:
\begin{itemize}
    \item Device 6 is less loaded than device 8. 
    \item The links used when choosing device 6 are more loaded than when using device 8. 
    \item The latency on the links used when choosing device 6 is higher than when using device 8. 
\end{itemize}

Next, the results for all the 25 runs in the baseline group are analyzed. The placement obtained by the two objectives differed in 11 runs. 
Table \ref{tab:results_25rep} presents a summary of the results obtained for the complete group. 
It shows, for example, that the request completion time, the overall energy, and the marginal energy is distributed in a similar way for both objectives over the 25 runs. The solver execution times however appears consistently higher for the Overall energy objective, which justifies further investigation. 

To sum up, the Overall energy metric seems to be able to leverage the time remaining to the deadline to pick edge devices located further away but with more interesting energy consumption characteristics. On the contrary, the Marginal energy metric is more conservative and chooses edge devices located closer and where the current utilization is higher. These observations confirm the intuition for the idea behind the different metrics (see Section \ref{sec:objectiveFunction}) which justifies a further study of the alternative metrics in Sections \ref{sec:influenceLoadLevel} to \ref{sec:influenceCharacteristics} below. 

\begin{table*}[]
    \centering
    \begin{tabular}{l|c|c|c|c|c|c|c|c|c|c|}
    \cline{2-11}
     & \multicolumn{5}{c|}{Request completion time (ms)} & \multicolumn{5}{c|}{Overall energy consumed (J)}  \\
        \cline{2-11}
         & Mean & St. dev. & 10th & Median& 90th & Mean & St. dev. & 10th &Median & 90th   \\
        \hline
        \multicolumn{1}{|l|}{Overall energy}  &79&4&75&\cellcolor{blue!25}78&85&4176&478&3772&\cellcolor{blue!25}4110&4973   \\
        \hline
        \multicolumn{1}{|l|}{Marginal energy}  &77&3&75&\cellcolor{blue!25}77&81&4269&518&3772&\cellcolor{blue!25}4192&5096  \\
        \hline
        \cline{2-11}
        & \multicolumn{5}{c|}{Marginal energy consumed (J)} & \multicolumn{5}{c|}{Solver execution time (ms)} \\
        \cline{2-11}
         & Mean & St. dev. & 10th & Median& 90th & Mean & St. dev. & 10th &Median & 90th   \\
        \hline
        \multicolumn{1}{|l|}{Overall energy}  &1656&32&1623&\cellcolor{blue!25}1643&1694&79&36&49&65&129   \\
        \hline
        \multicolumn{1}{|l|}{Marginal energy}  &1629&10&1620&\cellcolor{blue!25}1628&1643&58&11&47&58&69  \\
        \hline
    \end{tabular}
    \caption{Statistical analysis of the evaluation results for a given request placement in 25 random load scenarios.}
    \label{tab:results_25rep}
\end{table*}

\subsection{Impact of load level and heterogeneity} 
\label{sec:influenceLoadLevel}

The previous results are obtained with a system being loaded around half of its capacity, both for the edge devices and the edge links. In order to further understand the impact of each energy metric, we vary the overall load level (i.e. the mean of the Normal distribution from Section \ref{sec:loadScenario}) for edge devices. All the edge links are completely underloaded in order to isolate the compute effect. This corresponds to row 2 in Table \ref{tab:evaluation_groups}. 
Moreover, we also want to vary the heterogeneity of the load, i.e. whether the load is fixed for all edge devices (no heterogeneity) or very different, some edge devices with a low load and some having a high load (high heterogeneity). This corresponds to rows 3 and 4 in Table \ref{tab:evaluation_groups}.

\begin{figure*}
     \centering
     \begin{subfigure}[b]{0.3\textwidth}
         \centering
         \includegraphics[width=\textwidth]{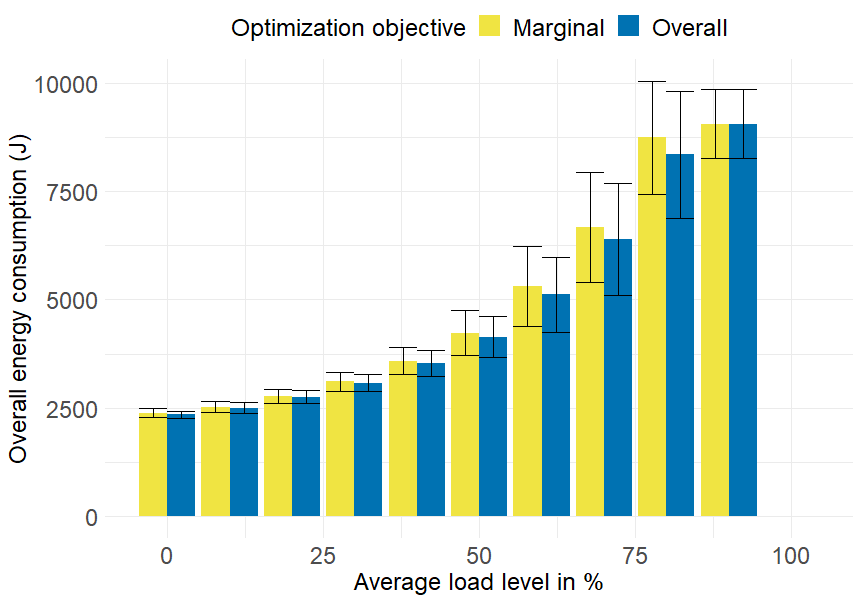}
         \caption{$\mathcal{N}$(load level, 10)}
         \label{fig:loadLevelHeterogeneous}
     \end{subfigure}
     \hfill
     \begin{subfigure}[b]{0.3\textwidth}
         \centering
         \includegraphics[width=\textwidth]{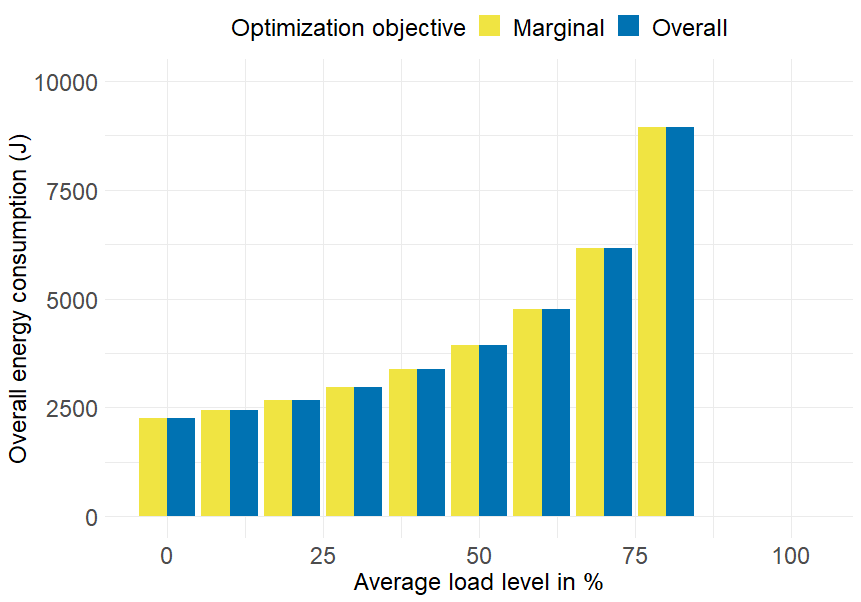}
         \caption{Fixed load at load level}
         \label{fig:loadLevelHomogeneous}
     \end{subfigure}
     \hfill
     \begin{subfigure}[b]{0.3\textwidth}
         \centering
         \includegraphics[width=\textwidth]{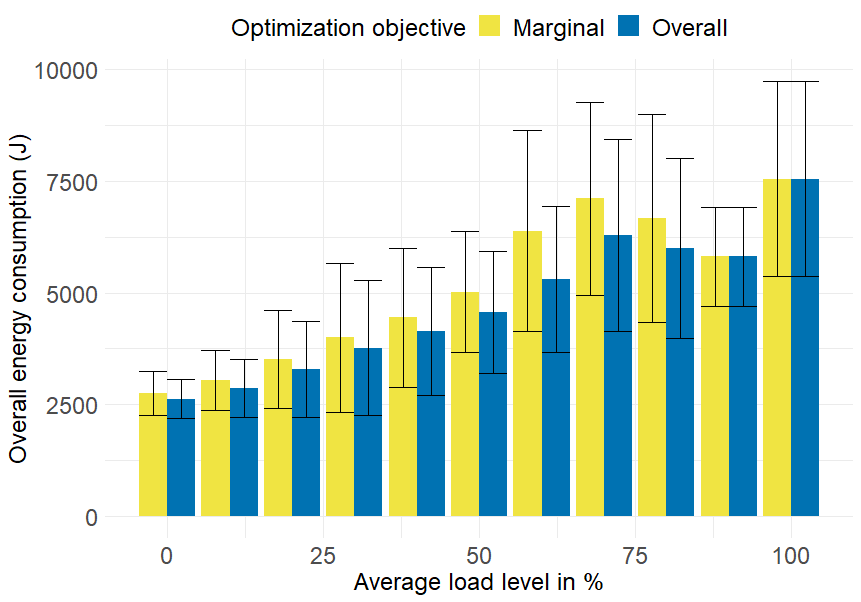}
         \caption{$\mathcal{N}$(load level, 30)}
         \label{fig:loadLevelHeterogeneous30}
     \end{subfigure}
        \caption{Average overall energy for a given request with different edge device loads. }
        \label{fig:loadLevels}
\end{figure*}

We start by studying the impact of different load levels, i.e. row 2 of Table \ref{tab:evaluation_groups}. The outcome after running 25 runs per load level is shown in Figure \ref{fig:loadLevelHeterogeneous}. This figure shows on the y-axis the overall energy resulting from one  request placement in different load conditions. The load varies in two ways: first, the x-axis consists of different load levels, each bar representing one load level (0 to 100\% in steps of 10) for a group of runs. Second, the load level is used as an input to the distribution which introduces different loads per run across the 25 runs. 

There are ten load levels (from 0 to 90\% average load) for which the solver is able to find a placement solution. The last one, where all the nodes are loaded at 100\% (i.e. no available resources) is infeasible for straightforward reasons 
(an infeasible scenario result in no data being shown). 
For a given load scenario, the two optimization objectives can result in the same placement or in a different one. This behaviour is similar for all load levels. The difference in placement decision thus does not seem to depend on the actual load level, i.e. whether the edge devices have a low load or a high load. 
Note that Figure \ref{fig:loadLevels} shows the overall energy on the y-axis, hence the value for the Marginal objective is always going to be higher or equal to the Overall one, per definition. If the placement solution obtained is the same, the two objectives result in the same value, otherwise they result in two different values. We show only the results for the overall energy for conciseness.  

On the other hand, we find out that the difference depends on the heterogeneity of the load levels in the edge infrastructure. Meaning that which optimization objective you choose will have an impact when you have (very) different load levels in the infrastructure. This is shown in Figures \ref{fig:loadLevelHomogeneous} and \ref{fig:loadLevelHeterogeneous30}, corresponding to rows 3 and 4 in Table \ref{tab:evaluation_groups}. 

In Figure \ref{fig:loadLevelHomogeneous}, the load is exactly the same on all edge devices, corresponding to the load level (this is row 3 in Table \ref{tab:evaluation_groups}). In that case, which objective is chosen does not matter, the placement will always be the same. This is shown by the bars in the figure being the same for the two objectives, and by the absence of standard deviation. Hence the absence of load heterogeneity leads to an absence of difference between the two optimization objectives. 

On the contrary, the differences between the edge device loads can be increased by increasing the standard deviation of the Normal distribution used from 10 to 30 (Figure \ref{fig:loadLevelHeterogeneous30}). This corresponds to a case where the available resources differ to a greater extent between the edge devices. In this case, the two optimization objective alternatives lead to different placements to a greater extent. This can be seen by larger standard deviations in Figure \ref{fig:loadLevelHeterogeneous30} than in Figure \ref{fig:loadLevelHeterogeneous}. 
 As an edge infrastructure in practice will exhibit high heterogeneity, this emphasizes the importance of carefully choosing the energy metric to optimize against. 

As we see by the highlighted cells in Table \ref{tab:results_25rep}, while a marginal energy optimization objective may on the face of it indicate a lower energy consumed (1628 vs 1643 J), it may actually lead to a higher overall energy consumption for the whole service chain (4192 compared to 4110 J) and the request completion time for both strategies is roughly the same (77 vs 78 ms).

\subsection{Impact of availability of function instances}
\label{sec:InfluenceAvailableInstances}

The next aspect to study is the availability of the function instances, i.e. the number of instances that are available for each function when a service request arrives. A higher number of function instances means more possible choices for the placement of the request and a lower one restricts the options of the optimization. 

We therefore reproduce the experiment presented on row 2 of Table \ref{tab:evaluation_groups}, which had 2 available function instances, with 4 available function instances (row 5) and 6 available function instances (row 6). 
The function instances are placed in a pre-defined way and so that two new instances of all the functions in the chain are added for each new experiment, but the already placed ones are not changed. 

\begin{figure*}
     \centering
     \begin{subfigure}[b]{0.3\textwidth}
         \centering
         \includegraphics[width=\textwidth]{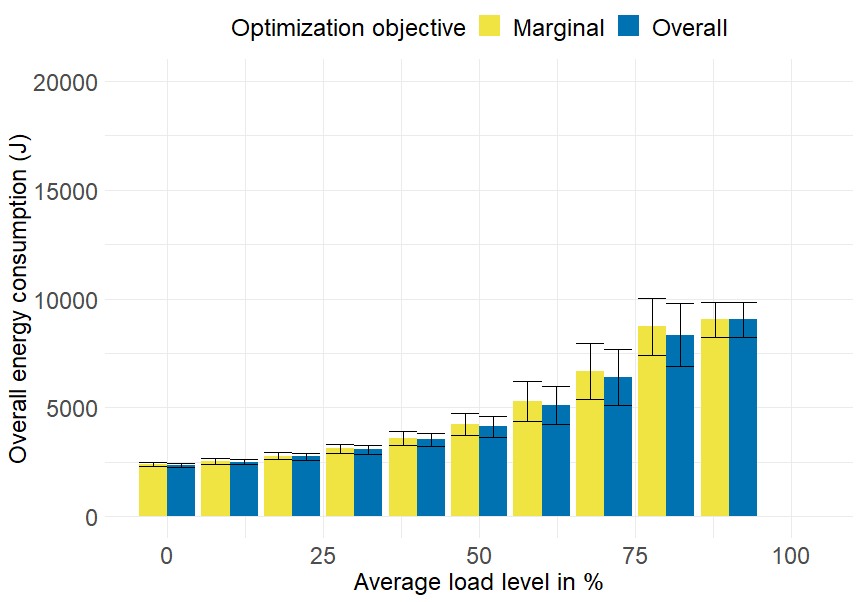}
         \caption{2 avail. function instances per function}
         \label{fig:functionInstance2}
     \end{subfigure}
     \hfill
     \begin{subfigure}[b]{0.3\textwidth}
         \centering
         \includegraphics[width=\textwidth]{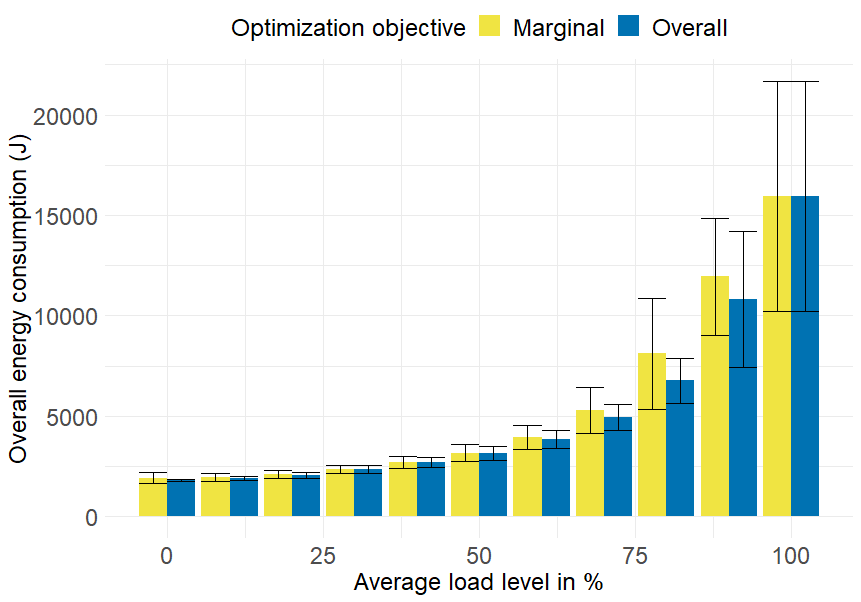}
         \caption{4 avail. function instances per function}
         \label{fig:functionInstance4}
     \end{subfigure}
     \hfill
     \begin{subfigure}[b]{0.3\textwidth}
         \centering
         \includegraphics[width=\textwidth]{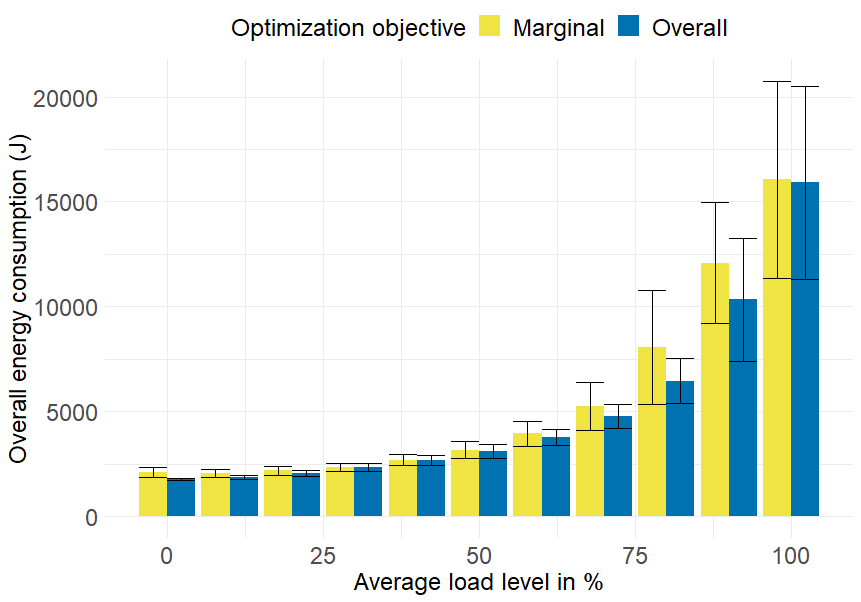}
         \caption{6 avail. function instances per function}
         \label{fig:functionInstance6}
     \end{subfigure}
        \caption{Average overall energy for a given request with varying function instance availability and infrastructure load levels.}
        \label{fig:functionInstances}
\end{figure*}

The outcomes are depicted in Figure \ref{fig:functionInstances}. As the number of function instances increases, moving from 2 to 6 instances per function (Figures \ref{fig:functionInstance2} to \ref{fig:functionInstance6}), the number of scenarios in which the different optimization objective alternatives lead to different placements increases. This is suggested by the figure data and confirmed using the logs. 

Also, the difference
in energy consumption between the alternatives (y axis in Figure \ref{fig:functionInstances}) presents differences at loads higher than 50\%.  Another interesting thing to notice was that the solver was unable to find a solution for a load level of 100 when only 2 function instances were available (no bar for 100 on Figure \ref{fig:functionInstance2}).  It was, however, possible when 4 function instances are available, and led to different placements depending on the objective when 6 functions were available. This illustrates how the solution space widens with more available function instances. 

In a nutshell, as the number of function instances increases, the number of load scenarios in which the different optimization objective alternatives lead to different placements increases. Moreover, the difference in energy consumption that appears between the objectives also increases. This further highlights the relevance of carefully choosing the energy metric used in the optimization objective. 

\subsection{Impact on request completion time}
\label{sec:influenceCharacteristics}

In Section \ref{sec:results}, we observed how the different optimization objectives led to different ways of placing the request. In particular, the Overall alternative seemed to favour devices located further away with lower load, while the Marginal alternative seemed to favour closer devices with current higher load, which is inline with the idea behind the different energy metrics. 
This flexibility of choosing different devices is possible as long as the completion time stays below the  deadline. 

\begin{figure*}
     \centering
     \begin{subfigure}[b]{0.3\textwidth}
         \centering
         \includegraphics[width=\textwidth]{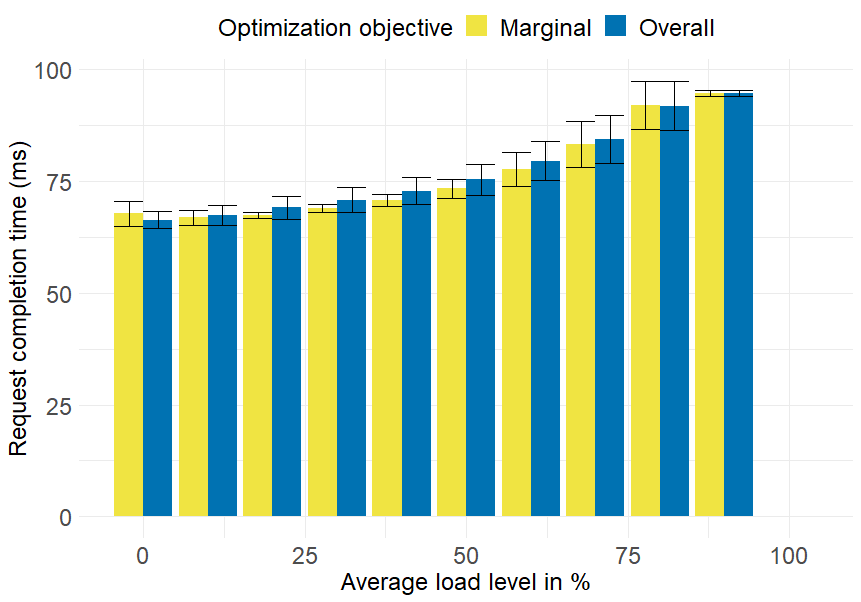}
         \caption{2 avail. function instances per function}
         \label{fig:completionTime2}
     \end{subfigure}
     \hfill
     \begin{subfigure}[b]{0.3\textwidth}
         \centering
         \includegraphics[width=\textwidth]{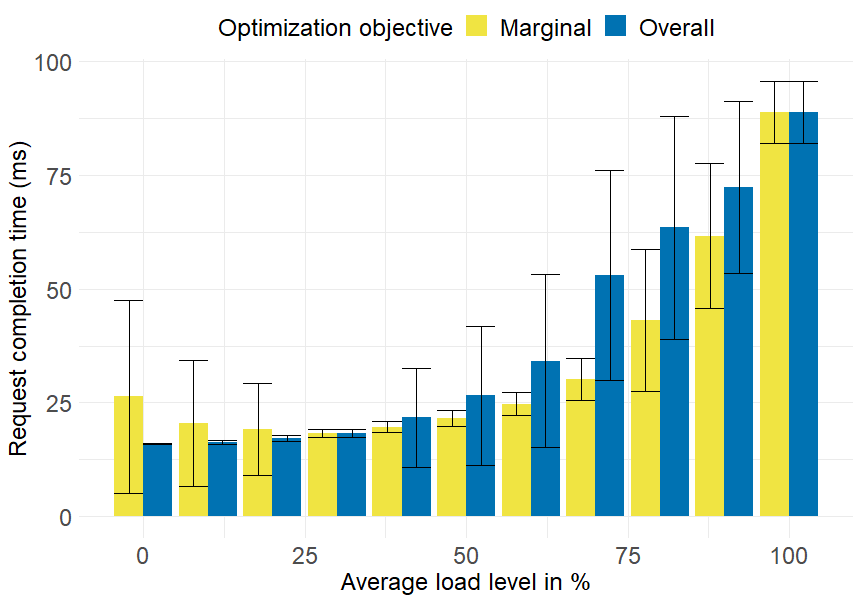}
         \caption{4 avail. function instances per function}
         \label{fig:completionTime4}
     \end{subfigure}
     \hfill
     \begin{subfigure}[b]{0.3\textwidth}
         \centering
         \includegraphics[width=\textwidth]{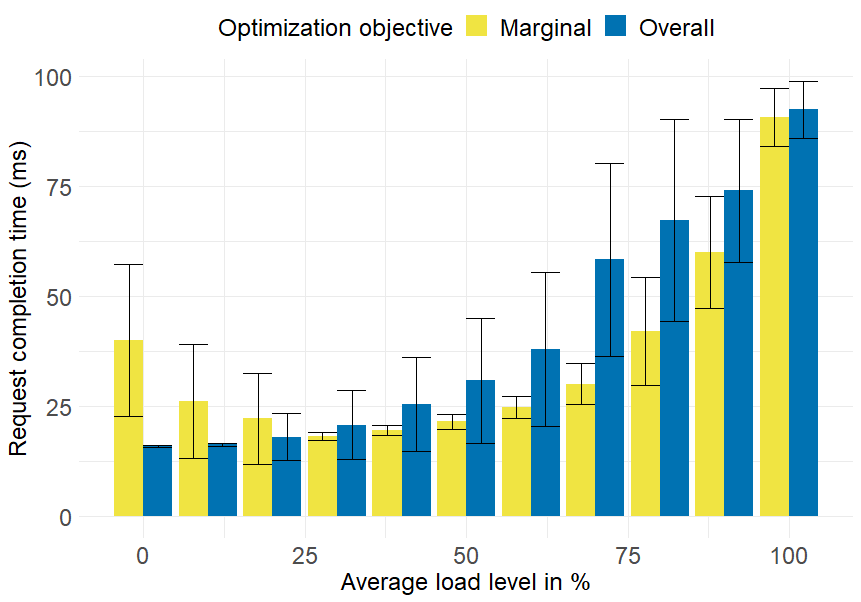}
         \caption{6 avail. function instances per function}
         \label{fig:completionTime6}
     \end{subfigure}
        \caption{Average completion time for a given request with varying function instance availability and infrastructure load levels.}
        \label{fig:completionTimes}
\end{figure*}

Here we look at the request completion time to see whether a lower overall energy is associated with a longer completion time, as a result of e.g. favoring devices located further away in the placement solution. Figure \ref{fig:completionTimes} shows the request completion time (on the y-axis) for the same runs as the one shown in Figure \ref{fig:functionInstances}, in order to analyze the trade-offs on completion time corresponding to the energy results presented in Section \ref{sec:InfluenceAvailableInstances}.

On average, the request completion time is higher for the Marginal optimization objective at low load levels and higher for the Overall optimization objective at higher load levels. This corresponds to the Overall metric indeed taking advantage of the time left to the deadline and trading off longer completion time (often a lot longer in the case of 4 or 6 function instances per function) for lower overall energy consumption. When the completion time is higher for the Marginal optimization objective, this is because the Marginal alternative avoids using an edge device that was previously unused, which is part of the idea behind choosing this alternative for optimizing.

\subsection{Time to solve the DRP problem}
\label{sec:eval_timeToSolve}
\begin{figure*}
     \centering
     \begin{subfigure}[b]{0.3\textwidth}
         \centering
         \includegraphics[width=\textwidth]{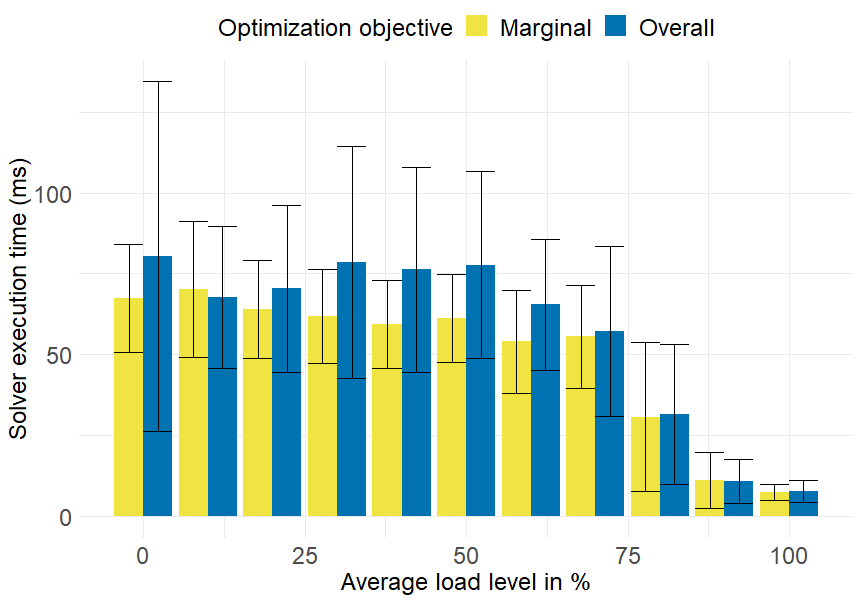}
         \caption{2 avail. function instances per function}
         \label{fig:executionTimes_2replicas}
     \end{subfigure}
     \hfill
     \begin{subfigure}[b]{0.3\textwidth}
         \centering
         \includegraphics[width=\textwidth]{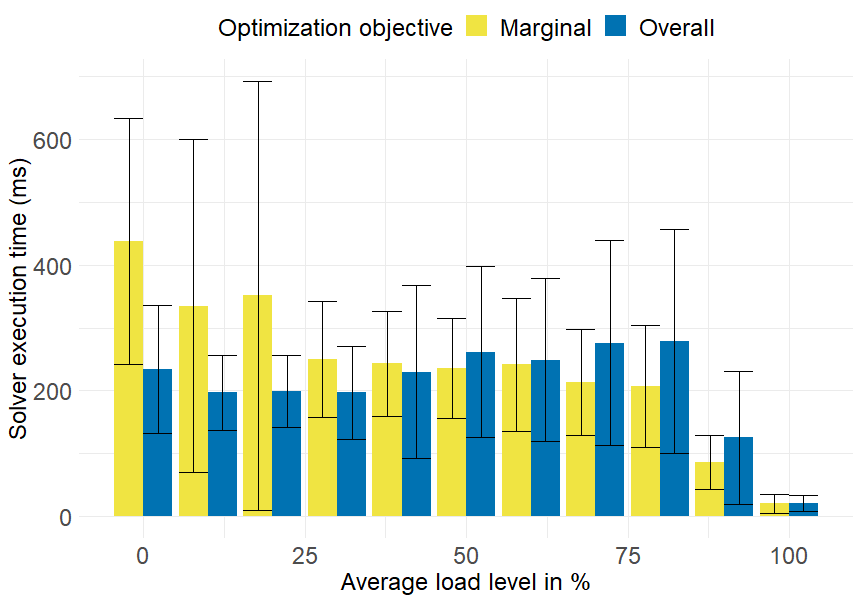}
         \caption{4 avail. function instances per function}
         \label{fig:executionTimes_4replicas}
     \end{subfigure}
     \hfill
     \begin{subfigure}[b]{0.3\textwidth}
         \centering
         \includegraphics[width=\textwidth]{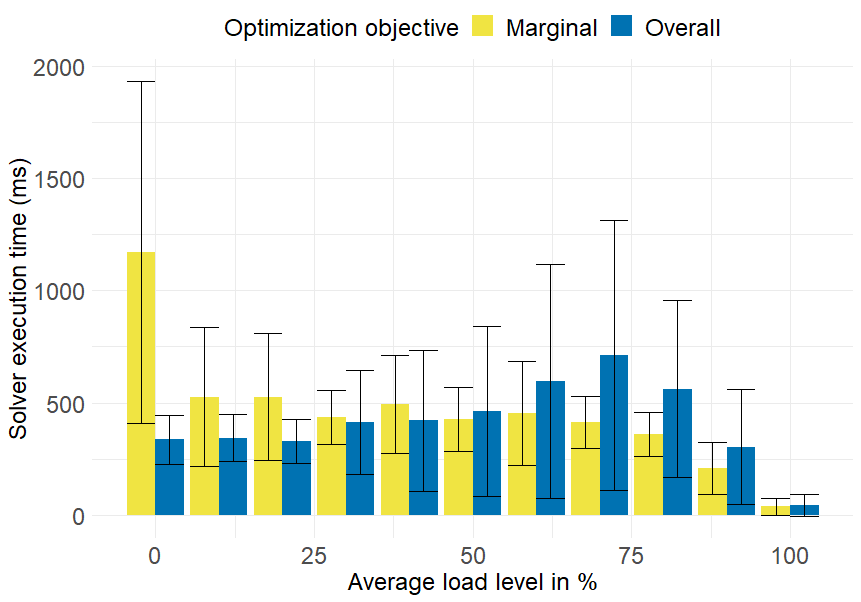}
         \caption{6 avail. function instances per function}
         \label{fig:executionTimes_6replicas}
     \end{subfigure}
        \caption{Solver execution time for a given request with varying number of function instances for every function. }
        \label{fig:executionTimes}
\end{figure*}

\
Finding a placement solution itself requires some time that is not included in the request completion time row/column of Tables \ref{tab:results} and \ref{tab:results_25rep} or in Figure \ref{fig:completionTimes}. 
The execution time of the DRP placement, i.e. the time needed for the solver to output a solution, is instead shown on a separate row/column in Tables \ref{tab:results} and \ref{tab:results_25rep} or in Figure \ref{fig:executionTimes}. 

As shown in Figure \ref{fig:executionTimes_2replicas}, using the Gurobi solver, a solution to the ILP problem from Section \ref{sec:results} (i.e. with two function instances per function) is found in between 34 ms and 180 ms. It can be seen that the solver execution time to find a solution does not depend on the load level, but significantly increases when the number of function instances increases (Figures \ref{fig:executionTimes_4replicas} and \ref{fig:executionTimes_6replicas}), as this widens the number of possible placements that have to be assessed. Moreover, which optimization objective leads to a solution fastest depends on the scenario (both with regards to load and function instance availability) and it does not seem possible to know in advance which one will be fastest. Finally, 
when the problem is infeasible, this is detected faster than when a placement solution can be found. 

Given the scale of the system and service evaluated, the computing capacity of the computer running the solver, 
and that the service considered has a deadline of 100 ms, such execution times are not suitable for a real-world placer. Therefore, we consider this work as a first step to understand the trade-offs involved in optimization of micro-services towards energy as a metric. As a next step, non-optimal alternatives that can deliver a faster execution time should be envisioned. 

The model can, however, be seen as a reference for the optimal placement, and also for understanding the impact of different energy metrics.

\section{Related Works}
\label{sec:RelatedWorks}

The microservice architecture~\cite{Cerny_MicroserviceArchitecture} is being extensively considered in the neighbouring field of cloud computing. It is a cornerstone of both the cloud-native~\cite{CloudNativeFoundation} and serverless~\cite{Li_ServerlessComputing_Survey} computing. 
Several surveys have presented the microservice works in cloud computing, including focusing on anomaly detection~\cite{Soldani_AnomalyDetection_Survey} or  practical dimensions such as Kubernetes scheduling~\cite{Carrion_KubernetesScheduling_Survey}. 
Microservices are also increasily used in the edge computing paradigm. The fine-grained modularity they offer is for example an asset for IoT applications~\cite{Pallewatta_PlacementMicroservicesIoT_Survey}. 

Regarding service request placement and as cloud computing is a centralized paradigm, most works have a centralized approach. For example, Shameli-Sendi et al. ~\cite{Shameli-Sendi_1_OptimalPlacement} also use a TPP formulation for simultaneously placing multiple service requests in a cloud environment. Contrary to this work, they use a centralized approach where requests for independent purchasers are placed simultaneously. They also do not consider energy nor that the request placement has to meet a deadline. In a later work~\cite{Shameli-Sendi_EfficientProvisioning}, they add a security dimension to the problem and also provide a heuristic to help scaling to larger systems (in terms of number of nodes included). 
In our work, instead of constraints relating to the network security defense patterns, we have the constraint that not all nodes have the desired function instance. Moreover, scaling in this work is treated differently from looking at the infrastructure size only. Indeed, for edge latency-critical applications, the number of nodes that are suitable is going to be limited. Hence, not all nodes are equivalent (as they cannot all host any micro-service). 

Schneider et al~\cite{Schneider_DistributedOnline} study placement problems at the edge and propose a solution for service coordination, which they define as combining the online scaling, placement, scheduling and routing of a service request. Their solution uses distributed deep reinforcement learning. The results indicate that a distributed placement solution performs better than an equivalent centralized one. In our work, the focus is on the request placement subproblem. Hence, the scaling part, which corresponds to the other two subproblems, is executed by another part of the system. This constrains the placement as new function instances cannot be added in the infrastructure to accommodate a request. Our core contribution is to consider energy in addition to 
performance in the placement solution. 

Wang et al.~\cite{Wang_MicroserviceOriented} study the placement of service instances and 
requests in the context of the Internet of Vehicles. They present a two-layer system including three algorithms to 1) place the service requests to devices, 2) enable different service requests to use the same function instance and 3) reduce the number of service instances when they are not needed anymore. Their optimization objective is to minimize service completion time and resource (CPU,  memory, and bandwidth) utilization. Contrary to our work, they place several service requests at the same time and can decide to start new function instances when placing the requests, which leads to all nodes having all services and the unused ones have to later be removed by the second layer. 
Our separation of the placement subproblems leads to understanding each layer separately, and specially with a focus on overall energy consumption. 

Russo Russo et al.~\cite{RussoRusso_QoSAwareOffloading} propose a microservice request placement which is composed of two levels. The first level is a simple heuristic algorithm that places every incoming request according to a probability distribution for the different placement options (on the current device, on a cloud device, on a neighboring edge device or discarding the request). The second level is optimizing the probabilities used in the first level at regular time intervals. The optimization is performed using linear programming considering monetary cost constraints and resource availability constraints. The aim is to maximize the requests satisfying their response time requirements.  Contrary to our work, the energy consumption is not considered. Moreover, the optimization has a different purpose as it is not directly performing the placement. 

In the edge computing paradigm, the DRP problem has been also tackled for services that are not using the microservice architecture. For example, Fahs and Pierre~\cite{Fahs_Proximity} present Proxy-mity, a traffic routing system handling monolithic requests. These can be seen as microservice with only one function in the service chain. The requests are distributed to a relevant service instance taking  account of both load on the instances and proximity, defined as the latency-wise distance.
Machine learning techniques can also be used, as for many other resource provisioning problems~\cite{LeDuc_MLResourceProvisioning_Survey} but their goal is not to show analytical optimality of a solution.

There are several works considering energy consumption within edge computing, e.g. \cite{Gnibga_LatencyEnergyCarbonAware,Hadjur_Beekeeping,Li_SustainableEdgeRenewable,Perin_SustainableEdgeComputing}. To evaluate the proposed methods and techniques, relevant energy models are required. 
Baccarelli et al.~\cite{Baccarelli_EcoMobiFog} propose models for the energy consumption of both devices and communication links. Their focus is on virtualized and multi-core devices in the context of 5G. They then use the proposed models for jointly optimizing the computing and network energy of multi-tier mobile-edge-cloud ecosystems. The aim is to decide in which tier a given function of a microservice chain should be executed. 
Similarly, Ahvar et al.~\cite{Ahvar_EstimatingEnergy} propose energy models for different types of so-called cloud-related architectures. It also includes models for both the devices and the communication links between them. Their aim is to compare different types of architectures (more or less distributed) with regards to their energy consumption. 
To the best of our knowledge, none studies microservice placement with energy in focus. 

\section{Conclusion}
\label{sec:Conclusion}

In this paper, we focus on solving the DRP problem. this means deciding how to place an incoming service request to the edge which uses the microservice architecture.  We try to answer the question: which devices should be selected in order to obtain satisfactory quality of service with a high regard to energy efficiency? Our approach expresses the DRP problem as an instance of the traveling purchaser problem and then solves it using an Integer Linear Programming formulation with two different objectives. 
Our formulations focus on a distributed solution 
minimizing energy consumption instead of solely aiming at the fastest possible response time. 
The study highlights the energy consumption against service completion time trade-offs for energy-aware placements following two different energy objectives. 

Our evaluation 
shows that optimizing for overall energy leads to placement decisions that leverage an increased request completion time where edge devices with interesting energy characteristics but located further away are utilized. On the contrary, optimizing for marginal energy favours a placement decision performing load consolidation. 
In some cases, for example if the load is very even in the system, the same placement can be obtained with both energy objectives. On the contrary, higher heterogeneity in the system load leads to a larger difference between the placements obtained using the two energy-based objectives.

Calculating the optimal solution is, however, impractical for near-real-time applications, as the solver execution time was at least several tens of milliseconds. Since edge applications may have response time requirements of tens of milliseconds or lower, the solver execution time should be at the millisecond level or lower. Obviously the Gurobi solver is not suitable to be used as part of a run-time system but indicates the optimal placement, that serves as a reference. Moreover, the impact of the different optimization objectives is going to stay the same regardless of the method used to obtain the (near-) optimal placement solution. 

While our work initiates the study of placement strategies for energy-based decisions, more work remains in the area. 
For example, a two-tier placement strategy combining the two optimization objectives based on the current system state would be interesting to develop. Moreover, it would be interesting to study how/when to re-assess a placement decision, as an edge system is a very dynamic environment.   Finally, alternative fast decision strategies for finding a good enough solution  should be looked into. Possible options include the use of heuristics, meta-heuristics or machine learning. 

\bibliographystyle{IEEEtran}
\bibliography{./sample-base.bib}

\end{document}